**Non-magnetic ground state in $A_2WCl_6$ (A = Cs, Rb, K): A face-centered cubic system of spin-orbit-entangled $J = 2$ states**

T. Takayama[1,2,3], K. Ishii[4], S. Bette[1], J. Nuss[1], Y. Matsumoto[1], K. Fürsich[1], M. Minola[1], D. P. Sari[5,6], I. Watanabe[6], A. Krajewska[1,2,7], R. Dinnebier[1], B. Keimer[1], and H. Takagi[1,2,3,8]

[1] Max Planck Institute for Solid State Research, Heisenbergstrasse 1, 70569 Stuttgart, Germany.

[2] Institute for Functional Matter and Quantum Technologies, University of Stuttgart, Pfaffenwaldring 57, 70569 Stuttgart, Germany.

[3] National Institute for Materials Science, 1-2-1 Sengen, Tsukuba 305-0047, Japan.

[4] Synchrotron Radiation Research Center, National Institutes for Quantum Science and Technology, Sayo, Hyogo 679-5148, Japan.

[5] Graduate School of Engineering and Science, Shibaura Institute of Technology, 307 Fukasaku, Minuma, Saitama 337-8570, Japan.

[6] Nuclear Structure Research Group, RIKEN Nishina Center for Accelerator-Based Science, Wako, Saitama 351-0198, Japan.

[7] ISIS Neutron and Muon Source, STFC Rutherford Appleton Laboratory, Chilton, Didcot, Oxon OX11 0QX, United Kingdom.

[8] Department of Physics, University of Tokyo, Bunkyo-ku, Tokyo 113-0033, Japan.

**Abstract**

Heavy transition metal compounds with strong spin-orbit coupling appeared as a platform for *d*-electron multipolar physics. We report the electronic, magnetic, and structural properties of antifluorite-type tungsten chloride $A_2WCl_6$ (A = Cs, Rb, and K), comprising a face-centered cubic lattice of $W^{4+}$ ions. The $5d^2$ configuration of $W^{4+}$ ions in a cubic environment yields a spin-orbit-entangled $J = 2$ state, which has been discussed to give rise to multipolar ordering such as charge quadrupolar or magnetic octupolar ordering. We found that $K_2WCl_6$ undergoes a cubic-to-tetragonal structural transition which lifts the degeneracy of the $J = 2$ state, leading to a non-magnetic singlet ground state. By contrast, $Rb_2WCl_6$ and $Cs_2WCl_6$ show no signs of phase transition and remain non-magnetic down to the lowest temperature measured. At low temperatures, signatures of weak structural anomalies were revealed, which may point to the presence of local distortions of the $WCl_6$ octahedra. We argue that the subtle structural distortion arises from the local quadrupolar component of the $J = 2$ state but the frustrated quadrupolar interaction, together with chemical disorder, inhibits the formation of long-range quadrupolar ordering.

## Introduction

The interplay between strong spin-orbit coupling (SOC) and electron correlation, particularly that in heavy transition-metal compounds containing 4*d* or 5*d* elements, has provided a plethora of novel quantum states of matter [1-3]. When 4*d*/5*d* elements are octahedrally coordinated by ligand anions, their *d*-electrons are accommodated in the $t_{2g}$ manifold due to the large cubic crystal field. The SOC of 4*d*/5*d* elements, which can be comparable to or even stronger than noncubic crystal field even if it is present, entangles the spin and orbital degrees of freedom, yielding a spin-orbit-entangled $J_{\mathrm{eff}}$ state. The magnetic interactions between localized $J_{\mathrm{eff}}$ pseudospins can take forms distinct from those between spin-only moments because of the inherited orbital moments [3,4]. Unprecedented magnetic ground states arising from these interactions, such as Kitaev spin liquid [5] and excitonic magnetism [6], have been theoretically predicted to emerge, and material realization has been intensively explored.

When the number of *d*-electrons in the $t_{2g}$ manifold is one or two, SOC forms $J_{\mathrm{eff}}$ = 3/2 or 2 state, respectively. These states have been proposed to display the ordering of electronic multipoles such as charge quadrupoles and magnetic octupoles [7-9]. The face-centered cubic (FCC) lattice has been put forward as a platform for such multipolar ordering, and rocksalt-ordered double-perovskite oxides with a general formula of $A_2B'B''O_6$ (A: alkali-earth or lanthanoid ions, B'(B"): transition-metal ion or other cations coordinated octahedrally by oxygen atoms) have been a particular focus in this respect. When one of the B cations, B", is non-magnetic, their magnetic and electronic properties are governed by the FCC sublattice of B' ions. In $5d^1$ double-perovskite oxides hosting $J_{\mathrm{eff}}$ = 3/2 state like $Ba_2NaOsO_6$ and $Ba_2MgReO_6$, a cubic-to-tetragonal structural transition has been observed, followed by long-range magnetic dipolar ordering [10-12]. The preceding structural transition, which releases sizeable electronic entropy, has been discussed to derive from the formation of charge quadrupole ordering out of the $J_{\mathrm{eff}}$ = 3/2 manifold [13,14].

The FCC lattice of $d^2$ ions was originally proposed to undergo analogous charge quadrupolar ordering from the $J_{\mathrm{eff}}$ = 2 manifold [8,9]. However, most of the $d^2$ double-perovskites have been reported to display only a single transition of long-range magnetic order or spin-glass freezing, and no charge quadrupole ordering has been identified [15,16]. The five-fold $J_{\mathrm{eff}}$ = 2 state in the cubic environment is expected to split into a doublet $E_g$ and a triplet $T_{2g}$ because of the residual cubic crystal field, i.e. by the admixture of excited $t_{2g}^1e_g^1$ configuration induced by spin-orbit coupling [3, 17,18] (we denote the *d*-orbital manifolds split by cubic crystal field as $t_{2g}$ and $e_g$ in italic lower case, while the doublet and triplet out of the $J_{\mathrm{eff}}$ = 2 state are shown as $E_g$ and $T_{2g}$ in upper case). The lower-energy $E_g$ state is a non-Kramers doublet carrying no magnetic dipole moment, but it does host a charge quadrupole and magnetic octupole moment, which may exhibit long-range ordering at low temperatures. Although the $E_g$-$T_{2g}$ split was formerly assumed to be smaller than the magnitude of exchange interactions and thus negligible [8], the presence of this gap with a size of 10-20 meV was experimentally confirmed by inelastic neutron scattering measurements on double-perovskite oxides with $5d^2$ $Os^{6+}$ ions [17]. Notably, the osmates have been proposed to develop magnetic octupolar ordering from the $E_g$ doublet. $Ba_2MgOsO_6$ displays a phase transition at ~50 K with a subtle anomaly in the magnetic susceptibility, accompanied by an entropy change close to $R\ln 2$. Although muon-spin precessions were observed in the zero-field condition below the transition temperature, no magnetic Bragg peak was found in the neutron diffraction. Furthermore, powder X-ray diffraction measurements did not reveal any structural distortions expected for a quadrupolar-ordered state. The formation of magnetic octupolar ordering has been advocated to fully account for these observations [17-19].

Nevertheless, the ground state of $d^2$ double perovskites remains under debate. The appearance of magnetic octupole ordering was initially attributed to the predominant ferro-octupolar coupling arising from interactions involving the excited $T_{2g}$ manifold [18]. Pourovskii et al. performed many-body first-principle calculations and showed that octupolar ordering is formed by the predominant ferro-octupolar interaction between the $E_g$ doublets [20]. In contrast, Khaliullin et al. argued from the microscopic theory that the quadrupole interaction is stronger than octupolar coupling when the dominant *d*-electron hopping processes are considered, and the formation of charge-quadrupole ordering should be more plausible [21]. Churchill and Kee extended the investigation by including all the allowed exchange processes and showed that the Os double-perovskites locate at the verge of magnetic octupole and charge quadrupole ordered states [22]. In contrast to the osmates, the cubic double-perovskite rhenate $Ba_2YReO_6$ with $Re^{5+}$ ($5d^2$) ions undergo long-range magnetic dipolar ordering, as shown by polarized neutron scattering measurements [23]. This is argued to arise from the lower-energy triplet with the reversed energy hierarchy of $E_g$-$T_{2g}$ states [19,24].

A large body of previous discussions on the ground state of $d^2$ double-perovskites suggests the presence of rich electronic phases in a subtle balance of competing interactions. That is, fine-tuning of relevant electronic parameters such as spin-orbit coupling, Hund's coupling and degree of *d-p* hybridization may result in a formation of distinct multipolar phases. We therefore visited the antifluorite $K_2PtCl_6$-type tungsten chloride $A_2WCl_6$ (A = Cs, Rb, K), which comprises an FCC lattice of $W^{4+}$ ($5d^2$) ions [25,26]. The $W^{4+}$ ion is octahedrally coordinated by $Cl^-$ ions in a perfect cubic environment. The reduced hybridization between $5d$ $W^{4+}$ and $Cl^-$ ions, compared to those in $5d$ oxides, may change the landscape of multipolar interactions. Previous studies on $A_2WCl_6$ [A = Cs, Rb] found no signs of magnetic ordering, although the nature of the magnetic ground state or electronic structure has not yet been clarified [27]. Our resonant inelastic X-ray scattering (RIXS) measurements indicate that the $5d^2$ electrons of $W^{4+}$ ions form a spin-orbit-entangled $J = 2$ state, which provides a platform for multipole physics. Nevertheless, $Cs_2WCl_6$ and $Rb_2WCl_6$ do not show a clear sign of phase transition down to the lowest accessible temperature, whereas $K_2WCl_6$ undergoes a structural transition to a tetragonal phase, likely induced by steric instability. No magnetic ordering was identified in $A_2WCl_6$ [A = Cs, Rb] in both thermodynamic and muon-spin relaxation (μSR) measurements. Although $A_2WCl_6$ [A = Cs, Rb] appears to retain cubic symmetry at low temperatures, subtle signatures of structural distortion were discerned in our low-temperature Raman scattering and powder X-ray diffraction (PXRD) measurements. We speculate that the individual $WCl_6$ octahedra are slightly distorted at low temperatures and the degeneracy of the $E_g$ doublet is lifted by quadrupolar interactions, leading to a non-magnetic ground state, but frustrated quadrupole coupling, in the presence of chemical disorder, hinders the formation of long-range ordering.

## Results

### Electronic structure and magnetic properties of $A_2WCl_6$ (A = Cs, Rb, K).

The obtained samples were found to be nearly a single phase of $A_2WCl_6$ with a small amount (1-2 wt%) of ACl impurity (see Supplementary Fig. S1 for the PXRD patterns). $A_2WCl_6$ crystallizes in a $K_2PtCl_6$-type cubic structure (space group: *Fm*-3*m*, No. 227) at room temperature, and the $WCl_6^{2-}$ octahedra and $A^+$ cations form an antifluorite-type arrangement [25,26], where $WCl_6$ octahedra are isolated from each other (Fig. 1**a**). The $W^{4+}$ ions are placed in a perfect cubic environment and compose the FCC sublattice. This structure can be viewed as a rocksalt-ordered double-perovskite structure with one of the B-site cations vacant. The $W^{4+}$ ions have two 5*d* electrons that are accommodated in the $t_{2g}$ manifold. The $S = 1$ and $L_{\text{eff}} = 1$ state of the $(t_{2g})^2$ configuration in the presence of spin-orbit coupling is expected to form a $J_{\text{eff}} = 2$ state in the *LS*-coupling scheme.

To clarify the local electronic state, we performed W $L_3$-edge RIXS measurements on single crystals of $Rb_2WCl_6$ and $K_2WCl_6$ and a polycrystalline pellet of $Cs_2WCl_6$. Figure 1**c** shows the obtained RIXS spectra measured with the incident X-ray energy $E_i$ of 10.203 keV. The spectra of $Cs_2WCl_6$ and $Rb_2WCl_6$ were collected at $T = 8$ K, and the data of $K_2WCl_6$ were taken at room temperature which is well above the structural transition temperature discussed below. In addition to the elastic signal, several electronic excitations were clearly resolved around 0.4, 1.0, 1.9, and 3.2 eV for all the samples. The observed peak did not show a visible momentum **Q**-dependence, indicative of the localized nature of the intra-atomic excitations. The peak intensities above 3 eV were enhanced when the energy of the incident X-rays was increased to 10.206 keV, which means that they represent the excitations between the $t_{2g}$ and $e_g$ manifolds of the W 5*d* orbitals, namely those across 10Dq (see Supplementary Fig. S2 for the $E_i$- and **Q**-dependences and Fig. S3 for the fitting of the RIXS spectra). The $t_{2g}$-$e_g$ excitation shows a shoulder on the high-energy side, and its origin is unclear at moment. We suspect that this may be ascribed to the spin-conserving and spin-flip processes or to the $J_{\mathrm{eff}} = 3/2 - 1/2$ split of the final $(t_{2g})^1(e_g)^1$ state, namely, $(J_{\mathrm{eff}} = 3/2)^1(e_g)^1$ and $(J_{\mathrm{eff}} = 1/2)^1(e_g)^1$ states.

The peaks below 2.5 eV represent excitations within the $t_{2g}$ manifold. We consider a single-ion Hamiltonian including SOC in the cubic limit, to clarify the relevant intra-atomic excitations. Figure 1**b** illustrates the schematic energy diagram of the spin-orbital multiplet for the $(t_{2g})^2$ configuration in the *LS* coupling scheme. The $(t_{2g})^2$ electronic configuration splits into $^3P$ ($S = 1$, $L_{\mathrm{eff}} = 1$), $^1D_2$($S = 0$, $L_{\mathrm{eff}} = 2$), and $^1S_0$ ($S = 0$, $L_{\mathrm{eff}} = 0$) states by Hund's coupling $J_{\mathrm{H}}$. By spin-orbit coupling $\zeta$, the $^3P$ state further splits into $^3P_2$ ($J_{\mathrm{eff}} = 2$), $^3P_1$($J_{\mathrm{eff}} = 1$), and $^3P_0$ ($J_{\mathrm{eff}} = 0$) states. Figure 1**d** shows the energy levels of these spin-orbital multiples as a function of $\zeta$, where $J_{\mathrm{H}}$ is fixed at 0.3 eV, a reasonable value for 5*d* compounds [28]. The *LS*-coupling picture is valid for $J_{\mathrm{H}} >> \zeta$, while the large $\zeta$ limit corresponds to the *jj*-coupling regime where two *d*-electrons are accommodated in the $j_{\mathrm{eff}} = 3/2$ manifold (Supplementary Fig. S4 shows the energy levels in the large $\zeta$ limit. Here, we denote the spin-orbital multiplets of *LS*-coupling scheme with upper case letters, whereas those of the *jj*-coupling picture are expressed by lower case). The vertical axis represents the energy difference from the ground state. The peak positions in the RIXS spectrum of $Rb_2WCl_6$ are shown as red horizontal lines in the figure. The peaks at 0.95 and 1.85 eV can be attributed to the excitations from the ground state to the $^1D_2$ and $^1S_0$ state, respectively. The one at 0.36 eV represents the excitation to the states corresponding to the $J_{\mathrm{eff}} = 1$ and 0 states in the small $\zeta$ limit. By optimizing $\zeta$ and $J_{\mathrm{H}}$ numerically to the peak positions, we obtained $\zeta \sim 305$ meV and $J_{\mathrm{H}} \sim 310$ meV, indicating that $A_2WCl_6$ locates in the intermediate region between the *LS*- and *jj*-coupling schemes, as in other 5*d* compounds [28-30]. Despite being in the intermediate regime, the ground state of the $5d^2$ configuration in $A_2WCl_6$ retains five-fold degeneracy, and we denote this ground state as the $J = 2$ state hereafter.

As discussed in Introduction, the five-fold $J = 2$ is expected to split into the $T_{2g}$ triplet and $E_g$ doublet by the cubic crystal field, but this was not resolved in the RIXS spectra. The energy gap $\Delta_c$ between $E_g$ and $T_{2g}$, which was found to be 10-20 meV for the Os double-perovskites [17], is proportional to $\zeta^2$/10Dq [18,21]. The $\zeta$ of W is close to that of Os, while the 10Dq of $A_2WCl_6$ (~3.2 eV), is slightly larger than that of Os double-perovskite oxides (4-5 eV). The possible $\Delta_c$ of $A_2WCl_6$ should be similar in magnitude to that of $Ba_2MgOsO_6$, that is, a few tens of meV. This energy scale is well below the energy resolution of current RIXS experiments, and the excitation, if any, could be masked by elastic scattering. Moreover, the $E_g$-$T_{2g}$ excitation was not identified in the powder inelastic neutron scattering (INS) spectra of $Rb_2WCl_6$ either (Supplementary Fig. S5). The INS map is dominated by strong phonon excitations below 50 meV, and no clear signs of magnetic excitations could be resolved.

Nevertheless, we cannot exclude the cases where the $E_g$-$T_{2g}$ excitation is present at a higher energy above 30 meV but decays quickly with $|Q|$ or it is strongly coupled with the phonon excitations and substantially broadened, making it difficult to resolve by powder INS.

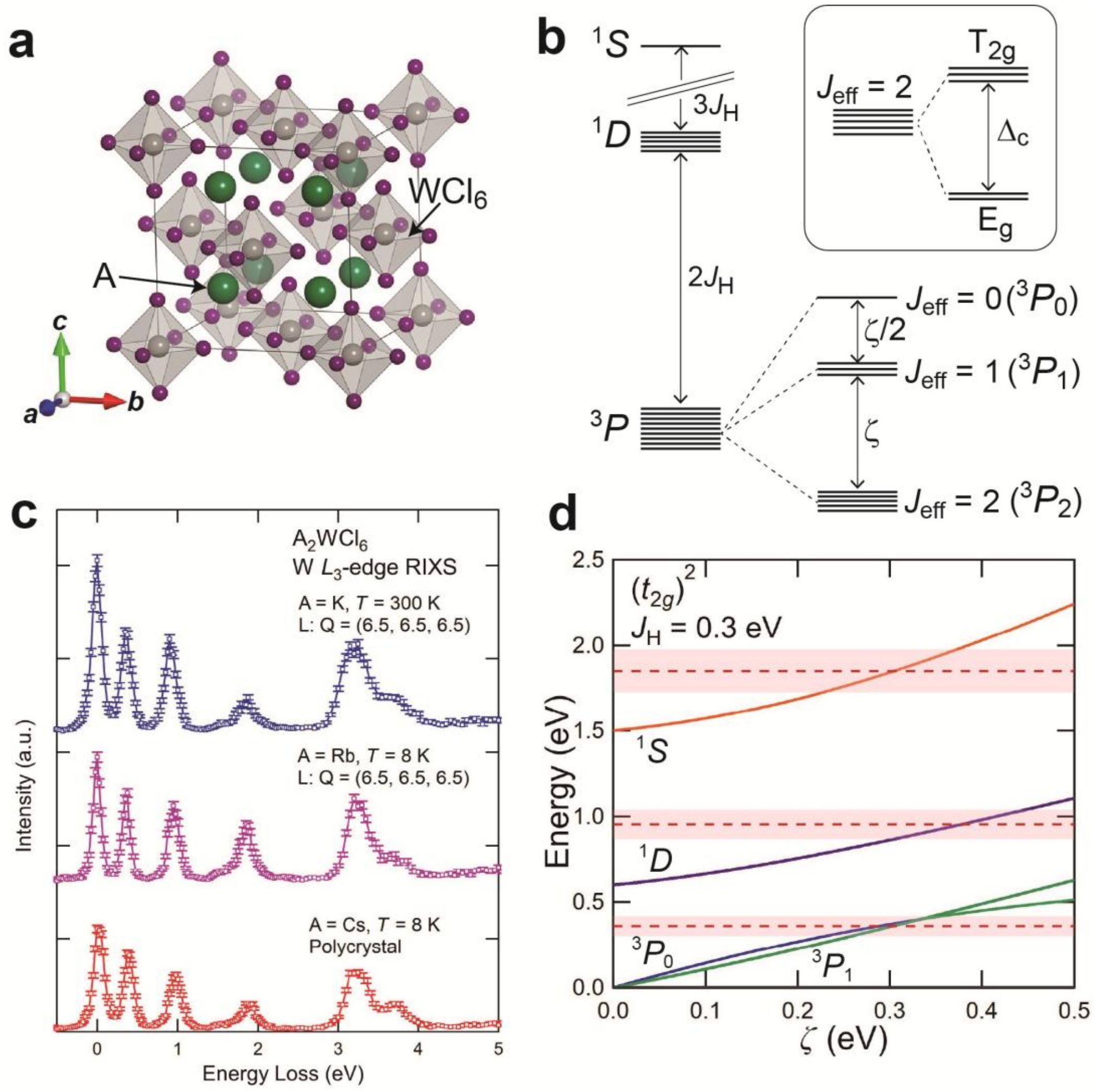

Fig. 1 **a**. Crystal structure of $A_2WCl_6$ (A = Cs, Rb, and K) forming a cubic lattice with the space group of *Fm*-3*m* (No. 225). **b**. Spin-orbital multiplet structure of $d^2$ configuration in the *LS*-coupling scheme. The inset depicts the split of $J_{\mathrm{eff}} = 2$ ($^3P_2$) state into the non-Kramers doublet $E_g$ and the $T_{2g}$ triplet with a gap of $\Delta_c$. **c**. W $L_3$-edge RIXS spectra for $A_2WCl_6$. The energy of incident X-ray is 10.203 keV. The measurements on $K_2WCl_6$ and $Rb_2WCl_6$ were performed on the single crystals, while the polycrystalline pellet was used for $Cs_2WCl_6$. The spectra for the formers were collected at the L point of the Brillouin zone. **d**. Calculated energy levels for spin-orbital multiplets of $(t_{2g})^2$ configuration as a function of spin-orbit coupling $\zeta$. The vertical axis shows the energy difference from the ground state. The Hund's coupling $J_H$ is fixed at 0.3 eV. The horizontal red dashed lines show the RIXS peak positions for $Rb_2WCl_6$, and the pink shaded areas correspond to the width of each excitation peak.

**Magnetic and thermodynamic properties of $A_2WCl_6$.**

The $J = 2$ state of $A_2WCl_6$ is expected to display transitions associated with the ordering of charge or magnetic multipoles. Figures 2**a** and **b** show the magnetic susceptibility $\chi(T)$ and heat capacity $C(T)$ measured for the polycrystalline samples. $\chi(T)$ shows a Curie-Weiss-like increase upon cooling from room temperature, but no clear anomalies are seen down to 2 K for all the $A_2WCl_6$ samples. The effective moment $\mu_{eff}$ and Curie-Weiss temperature $\theta_{CW}$ are 1.41 $\mu_B$/W and -62 K for $K_2WCl_6$, 1.46 $\mu_B$/W and -51 K for $Rb_2WCl_6$, and 1.39 $\mu_B$/W and -46 K for $Cs_2WCl_6$, respectively, as estimated from the Curie-Weiss fit shown in the inset. The magnitude of $\mu_{eff}$ is close to the value expected for the $J_{eff}$ = 2 moment with a $g$-factor of 1/2 [3], that is, 1.22 $\mu_B$. The deviation of $\mu_{eff}$ from that of an ideal $J_{eff}$ = 2 is smaller in $A_2WCl_6$ than in the $5d^2$ double-perovskite oxides ($\mu_{eff}$ = 1.5-2.3 $\mu_B$), most likely reflecting the reduction of $d$-$p$ hybridization that otherwise suppresses the orbital moment. Effective magnetic moments of nearly isolated paramagnetic ions have been well described by the Kotani model [31], while some deviations from the model have been found for $A_2WCl_6$ [27]. Our data of $Rb_2WCl_6$ deviates from the Kotani model as well, especially at low temperatures (Supplementary Fig. S6**a**), which may be attributed to the presence of inter-site interactions [32] or the formation of a non-Kramers $E_g$ doublet [33].

The heat capacity $C(T)$ of $Cs_2WCl_6$ and $Rb_2WCl_6$ did not show any signature of a phase transition down to 2 K as shown in Fig. 2**b**. No phase transition was observed down to 50 mK either in the measurements of the $Rb_2WCl_6$ single crystals (Fig. 3**a**). Below 1 K, a broad hump of $C(T)/T$, centered at ~0.2 K in zero field, was observed, and the hump shifted to higher temperatures when a magnetic field was applied. This suggests that the hump originates from defect spins. The entropy change associated with the defect spins was estimated to be approximately 0.9 J/mol·K up to 2 K, which is about 15% of spin-1/2. Similar amounts of defects were also found in the magnetization curve at low temperatures, where the Brillouin function-like contributions, superposing on the paramagnetic background, were attributed to the defect spins (Supplementary Fig. S6**b**). Such a sizeable number of defects may be attributed to the presence of chemical or structural disorders, as discussed below. In contrast, the $C(T)$ of $K_2WCl_6$ clearly indicates the appearance of a phase transition at 180 K (Fig. 2**b**). As discussed below, this transition represents a cubic-to-tetragonal structural transition. The entropy change estimated to be ~1.3 J/mol·K through this transition suggests the involvement of electronic degrees of freedom.

To obtain further insights into the ground state, μSR measurements in the zero-field condition (ZF-μSR) were performed on powder samples of $A_2WCl_6$ (A = K, Rb, Cs). Figs. 4**a** and **b** show the time dependence of the asymmetry parameter of the muon-spin polarization, A($t$), at the selected temperatures for $Rb_2WCl_6$ and $K_2WCl_6$, respectively. The data for $Cs_2WCl_6$ are presented in Supplementary Fig. S7, and the data in the presence of the longitudinal field of 100 Oe are shown in Fig. S8. For all the samples, A($t$) showed no oscillatory signals or the recovery of the so-called "1/3-tail" down to the base temperature of 2 K, indicating the absence of static magnetism such as long-range magnetic order or a spin-glass state. The absence of muon-spin precession contrasts with the Os double-perovskites in which the oscillatory response of ZF-μSR has been interpreted as the appearance of magnetic octupole ordering [17, 34]. A($t$) was fitted using a Lorentzian function $A(t) = A_0\exp(-\lambda/t) + A_{bg}$, where $A_0$ denotes the initial asymmetry at $t = 0$, and $\lambda$ represents the muon-spin relaxation rate. $A_{bg}$ is the background components associated with muons stopping at a silver sample holder and is assumed temperature-independent. The spectra exhibited only weak relaxation within the accessible experimental time window, and a simple exponential function was therefore adopted to characterize the temperature dependence of the relaxation rate. The temperature dependence of $\lambda$ is shown in Figs. 4**c** and **d**. $\lambda(T)$ increases monotonically upon cooling from room temperature in both samples. For $K_2WCl_6$, a small drop in $\lambda(T)$ is seen around 180 K where a phase transition was found in $C(T)$. Although no clear anomaly was seen in $\lambda(T)$ of $Rb_2WCl_6$ (Fig. 4**c**) and $Cs_2WCl_6$ (Supplementary Fig. S7), $\lambda(T)$ appears to level off around 100 K and then increases again upon further cooling. This behavior may be related to structural instability discussed in the next section.

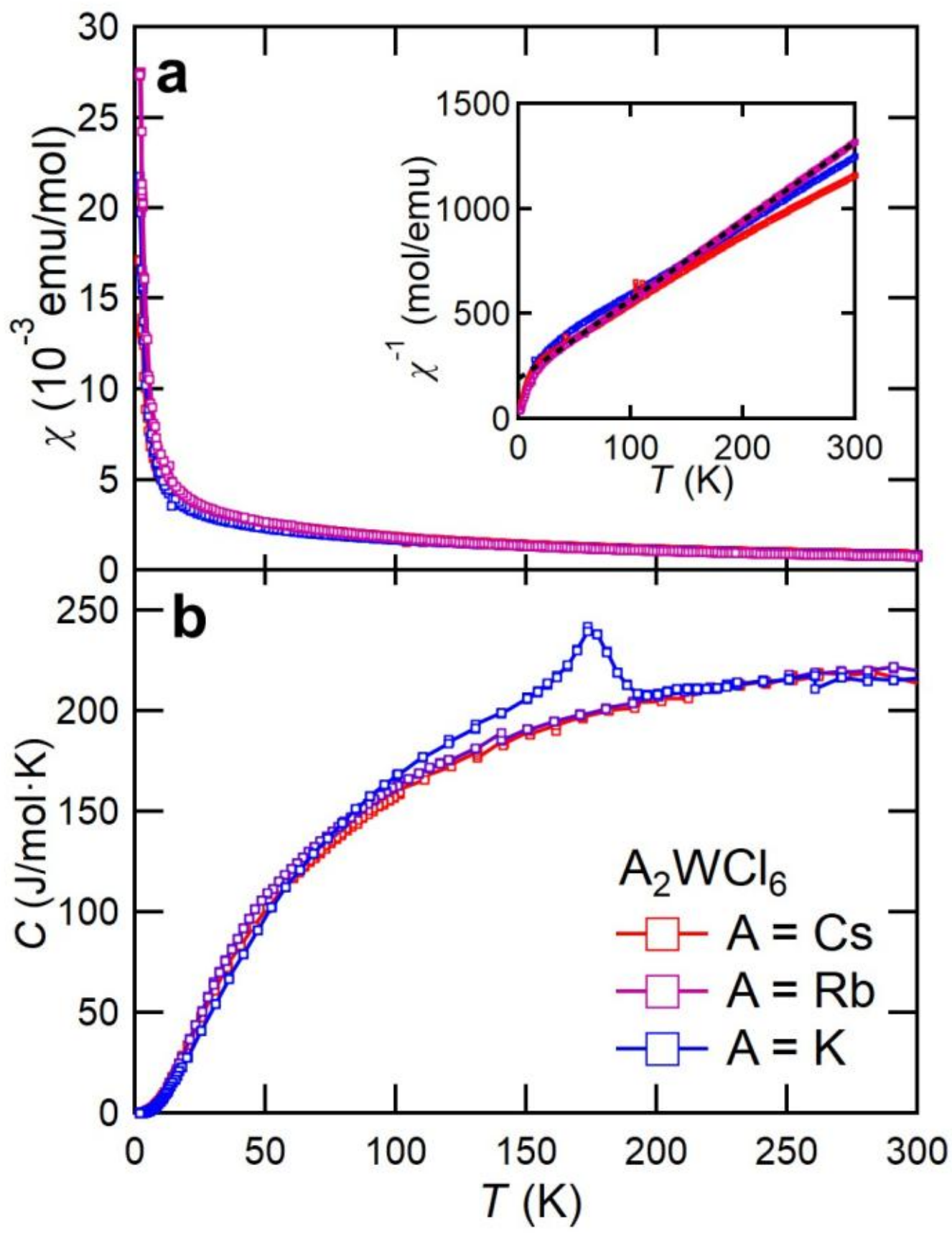

Fig. 2 Temperature dependence of **a**. magnetic susceptibility $\chi(T)$ and **b**. specific heat $C(T)$ of the polycrystalline samples of $A_2WCl_6$ (A = Cs, Rb, K). The inset of **a** shows the inverse of $\chi(T)$, and the black dotted line delineates the Curie-Weiss fitting for the data of $Rb_2WCl_6$.

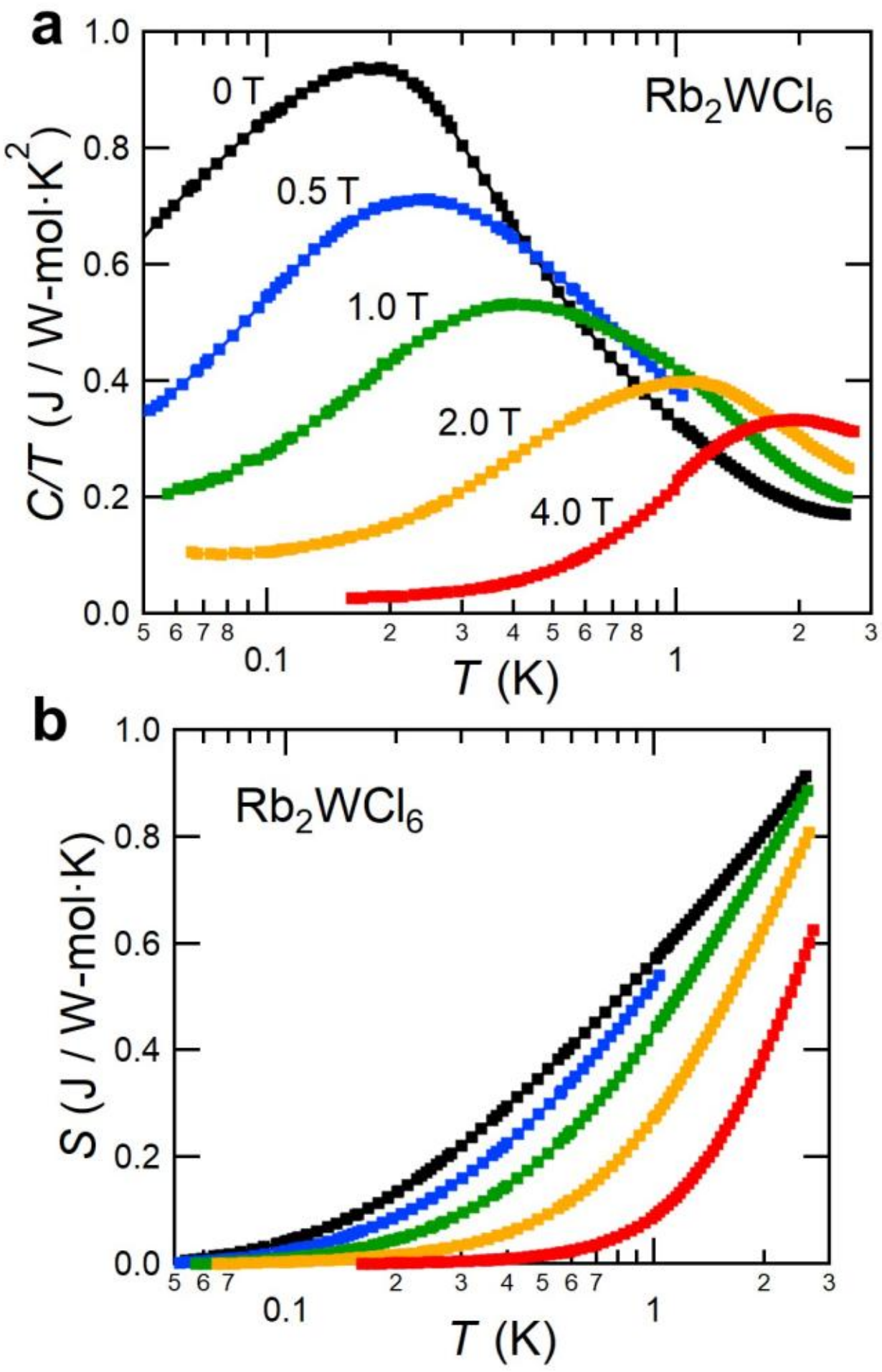

Fig. 3. **a**. Specific heat divided by temperature, $C(T)/T$, of $Rb_2WCl_6$ single crystals at low temperatures below 2 K, measured at several magnetic fields. The magnetic field is applied parallel to the [111] direction. **b**. Entropy change $S$ estimated from the $C(T)/T$ data. $S$ is as large as 0.9 J/W-mol·K at 2 K at zero magnetic field.

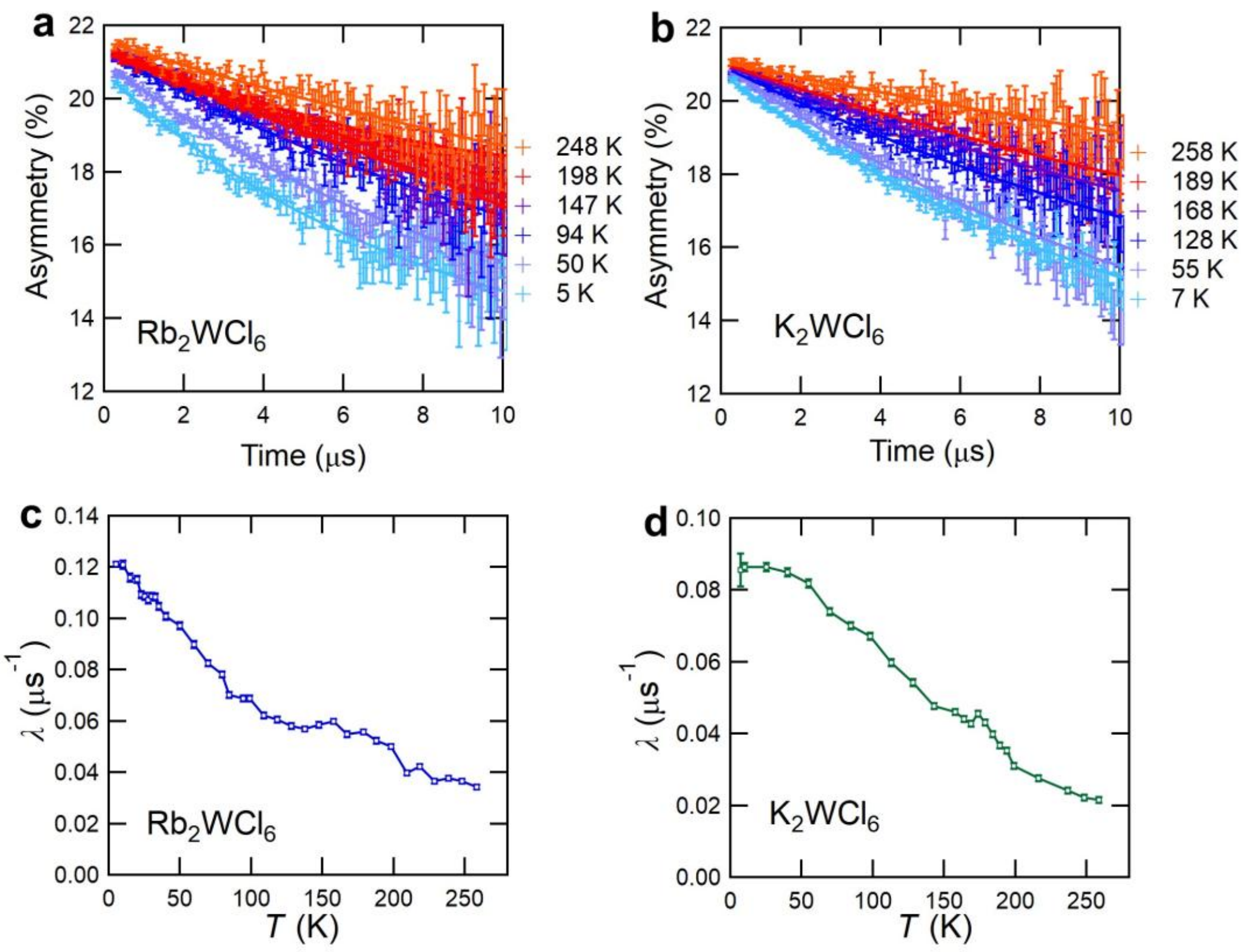

Fig. 4. **a**,**b**. Temperature dependence of μSR time spectrum measured on the powder of $Rb_2WCl_6$ and $K_2WCl_6$. The solid lines show the fit of experimental data by the Lorentzian function. **c**, **d**. Temperature dependence of muon-spin relaxation rate λ of $Rb_2WCl_6$ and $K_2WCl_6$.

**Structural instability of $Rb_2WCl_6$.**

The thermodynamic data presented above did not reveal any signature of phase transitions in $Rb_2WCl_6$ down to 50 mK. The ZF-μSR measurement excludes the presence of magnetic dipolar or octupolar ordering in contrast to the double perovskite oxides like $Ba_2YReO_6$ and $Ba_2MgOsO_6$ [16,17,23]. Likewise, a spin-glass freezing of $J = 2$ moments is ruled out. The ground state of $Rb_2WCl_6$ thus appears to be non-magnetic.

One of the possible mechanisms leading to a non-magnetic ground state is the appearance of charge quadrupole ordering from the $J = 2$ manifold [8] or out of the $E_g$ doublet [18, 21, 22]. To explore this point, we investigated the low-temperature crystal structure of $Rb_2WCl_6$ using Raman spectroscopy and X-ray diffraction measurements. Figure 5**a** shows the Raman scattering spectra measured on the (111) surface of a $Rb_2WCl_6$ single crystal. The upper figure shows the spectra obtained for the parallel-geometry where the incident and scattered light have the same polarization, while the lower one shows the data from the crossed-geometry with orthogonal polarizations. The polarization direction of the incident light was arbitrary within the (111) surface. In the crystal structure of $Rb_2WCl_6$ (space group $Fm$-3$m$), there are four Raman-active modes such as $A_{1g}$, $E_g$, and two $T_{2g}$ modes (to distinguish from the spin-orbital manifold, phonon modes are expressed in italic upper case). Three phonon peaks were clearly observed in the spectra. Considering the Raman spectra of other antifluorite-type $A_2MCl_6$ compounds (M: transition-metal), the three peaks at ~170, ~230, and ~350 $cm^{-1}$ can be assigned as $T_{2g}$, $E_g$, and $A_{1g}$ modes, respectively, which originate from the bending, elongation and breathing distortions of $WCl_6$ octahedra, respectively [27,35,36]. The three phonon modes did not show any clear changes between 300 and 15 K in both the parallel and crossed geometries, suggesting that $Rb_2WCl_6$ retains cubic symmetry. Nevertheless, a weak signature of a structural anomaly was found in the temperature dependence of the Raman spectra. Figure 5**b** shows a contour plot of the parallel-geometry data. Although no splitting of phonon peaks or appearance of new peaks is seen, the $T_{2g}$ phonon mode near 170 $cm^{-1}$ appears to broaden slightly upon cooling. The upper inset of Fig. 5**a** depicts the full width at half maximum (FWHM) for the $T_{2g}$ phonon mode in the parallel geometry, obtained by fitting the peak with a pseudo-Voigt function. The FWHM decreases on cooling from 300 K, but starts to increase below 150 K. A similar broadening was observed for the crossed geometry (lower inset of Fig. 5**a**). In contrast, the FWHM of the other two phonon modes decrease monotonically upon cooling (Supplementary Fig. S9). Such broadening of phonon peaks may suggest the presence of local distortion despite the absence of global structural transitions down to the base temperature.

In agreement with the findings from Raman spectra, X-ray diffraction measurements displayed signatures of local distortion. Single crystal X-ray diffraction (SCXRD) performed at 100 K indicates a cubic structure with the space group *Fm*-3*m* which is unchanged from room temperature (Supplementary Table S2). The PXRD diffraction patterns also showed no clear signs of structural transition down to 12 K as shown in Fig. 6**a**. Nevertheless, the temperature dependence of the cubic lattice parameter *a* is non-monotonic and shows a kink around 100 K as displayed in the inset of Fig. 6**a**, implying the presence of minute structural changes. Additionally, we found that some Bragg peaks such as (400) broadened at low temperatures as depicted in Fig. 6**b**. The FWHM of this peak increased gradually upon cooling (Fig. 6**c**). For other Bragg reflections, the (220) peak showed a weaker broadening at very low temperatures (Supplementary Fig. S10), while the (111) reflection did not show broadening upon cooling. This implies that specific local distortions grow in the lattice at low temperatures, although they are not coherent and do not induce a structural transition. Similar behaviors were found in $Cs_2WCl_6$ although the degree of broadening was weaker (Supplementary Fig. S11).

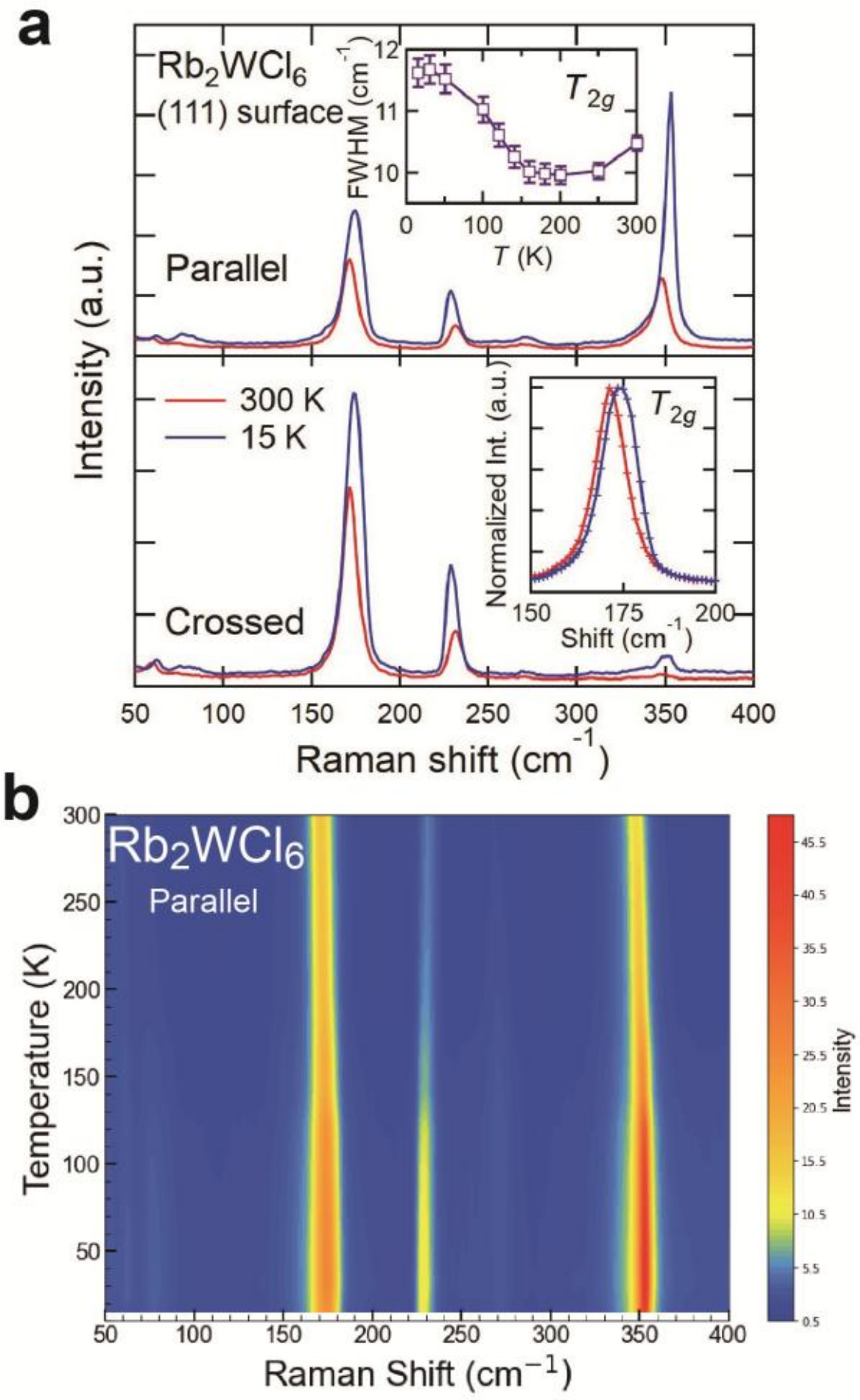

Fig. 5. Raman scattering measurement on a $Rb_2WCl_6$ single crystal. **a**. Raman scattering spectra in the parallel (upper) and the crossed geometries (lower) of light polarizations. The measurements were performed on the (111) surface of the single crystal, and the polarization of incident light was in an arbitrary direction withing the (111) surface. The upper inset displays the temperature dependence of full-width at the half maximum (FWHM) of the $T_{2g}$ phonon mode at ~170 cm$^{-1}$ in the parallel geometry. The lower inset shows the normalized intensities of the $T_{2g}$ phonon mode at 15 and 300 K in the crossed geometry. A slight broadening is seen at 15 K. **b**. Temperature dependence of Raman scattering spectra in the parallel geometry displayed as a contour plot.

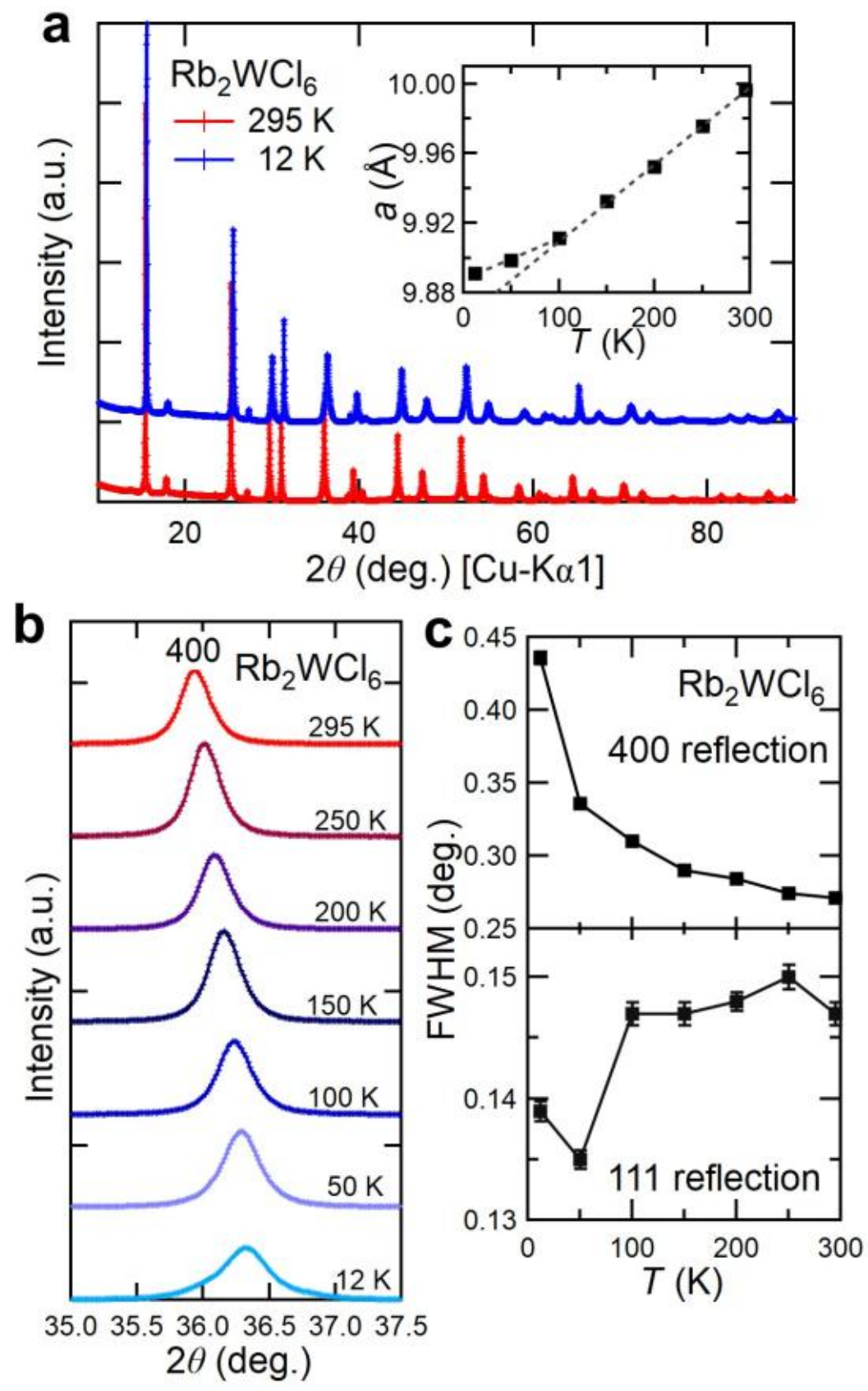

Fig. 6. Powder X-ray diffraction (PXRD) measurement on $Rb_2WCl_6$ at low temperatures. **a**. PXRD patterns at 12 and 295 K collected with Cu-Kα1 radiation. The data at 12 K is shifted vertically for brevity. There is no clear structural change down to 12 K. The inset depicts the temperature dependence of the cubic lattice parameter *a*. The two dashed lines show the linear fit above and below 100 K. A chang of slope is seen. **b**. (400) Bragg reflection of the PXRD patterns at selected temperatures. Upon cooling, some broadening of the peak is seen. **c**. Temperature dependences of full-width at half maximum (FWHM) for the (400) (upper) and (111) (lower) Bragg reflections. The increase of FWHM is seen for (400) reflection on cooling from room temperature, while no broadening is found for the (111) peak.

**Singlet ground state of $K_2WCl_6$ induced by structural transition**.

In contrast to $Cs_2WCl_6$ and $Rb_2WCl_6$, $K_2WCl_6$ displayed a clear sign of a phase transition around 180 K in the specific heat $C(T)$. SCXRD and Raman scattering indicate that this transition involves a structural change from the cubic to tetragonal phase. Figure 7**a** shows the crystal structure of $K_2WCl_6$ at 100 K from the SCXRD. The refined structural parameters are presented in Table 1. The $WCl_6$ octahedra are elongated along the $c$-axis and rotated about the $c$-axis. The bond length between the W and apical Cl atoms (W-$Cl_{ap}$) along the $c$-axis is 2.409(4) Å, whereas that in the $ab$-planes (W-$Cl_{eq}$) is 2.365(5) Å. The direction of rotation could not be refined satisfactorily, likely because of the presence of disorder associated with defects. The Raman spectra show a clear splitting of the $T_{2g}$ phonon mode near 170 $cm^{-1}$, and some new features are seen at 15 K, pointing to a symmetry lowering (Fig. 7**b**). The splitting of the $T_{2g}$ phonon mode occurred at the transition temperature of 180 K and grows gradually upon cooling (Fig. 7**c**).

This structural transition is accompanied by a change in the electronic ground state. The W $L_3$-edge RIXS spectrum at 8 K shows a sizeable spectral weight in the low-energy region adjacent to the elastic scattering (Fig. 8**a** and its inset). This contribution can be explained by the presence of low-energy excitation at ~80 meV superposed on the elastic scattering (the red and black dotted lines in Fig. 8**a**, respectively), pointing to the split of the $J$ = 2 manifold by the tetragonal distortion of the $WCl_6$ octahedra. Figure 8**b** shows the calculated energy levels of spin-orbital multiplets as a function of the tetragonal crystal field. We employed SOC $\zeta$ = 0.3 eV and Hund's coupling $J_H$ = 0.3 eV based on the results for $Rb_2WCl_6$. By increasing the tetragonal elongation distortion, the $J$ = 2 manifold splits into the upper two doublets ($J^z$ = ±2 and ±1) and the lower singlet ($J^z$ = 0). The excitation at around 80 meV can be accounted for by the excitation from the ground-state singlet to the $J^z$ = ±2 doublet when the tetragonal crystal field is about 130 meV. This indicates that the ground state of $K_2WCl_6$ is a non-magnetic singlet split from the $J$ = 2 manifold. An analogous low-energy excitation has been observed in the distorted double-perovskite $Sr_2YReO_6$ where a $Re^{5+}$ ion has a $5d^2$ configuration and the $ReO_6$ octahedra are distorted already at room temperature in the way that they are elongated in the $ab$-plane (see Supplementary Fig. S12). $Sr_2YReO_6$ shows no magnetic transition down to low temperatures [37], consistent with the singlet ground state. We speculate that the finite magnetic susceptibility of $K_2WCl_6$ at low temperatures, as well as that of $Sr_2YReO_6$ [37], could be ascribed to van Vleck paramagnetism associated with the excitation from the $J^z$ = 0 to $J^z$ = ±1 and ±2.

The structural transition thus results in a non-magnetic ground state in $K_2WCl_6$ by lifting the degeneracy of $J$ = 2 state. We suspect that this structural transition does not represent electronically-induced quadrupolar ordering. In fact, a similar structural transition has been observed in a number of $A_2MX_6$ compounds (X: halogen atom) when the A-site cation has a small ionic radius like $K^+$, regardless of whether M cations are magnetic or non-magnetic [35,36,38,39]. For instance, $K_2TaCl_6$ displays a structural transition yielding elongated $TaCl_6$ octahedra, in contrast to $Rb_2TaCl_6$ and $Cs_2TaCl_6$ where the tetragonal compression of $TaCl_6$ octahedra observed at low temperatures is attributed to the charge quadrupole ordering [40]. The structural instability of $K_2TaCl_6$ is associated with the size mismatch between the small $K^+$ ions and $TaCl_6$ octahedra. In line with the Ta chlorides, we speculate that the phase transition observed in $K_2WCl_6$ is not of electronic origin but is induced by steric instability with small $K^+$ ions.

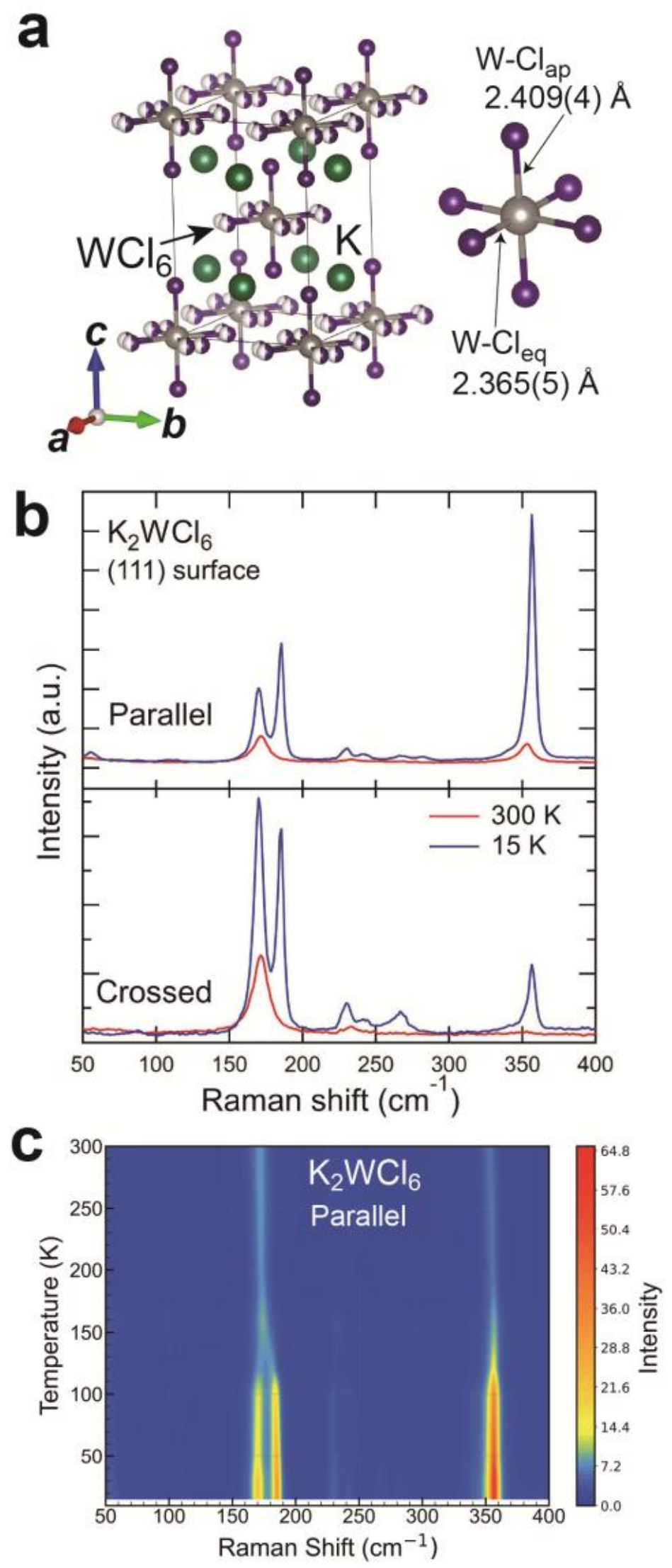

Fig. 7. Low-temperature crystal structure of $K_2WCl_6$. **a**. Crystal structure of $K_2WCl_6$ at 100 K analyzed by single crystal X-ray diffraction measurement. The $WCl_6$ octahedra are elongated along the *c*-axis and rotated within the *ab*-planes. The periodic pattern of the $WCl_6$ rotations was not determined precisely by the refinements, probably due to the presence of multiple domains associated with defect. The inset shows the close-up view of $WCl_6$ octahedra, showing the W-Cl bond lengths with the apical ($Cl_{ap}$) and with the equatorial ($Cl_{eq}$) chloride ions. **b**. Raman scattering measurements on the (111) surface of $K_2WCl_6$ crystal. The split of phonon modes, together with the appearance of new peaks, indicates the symmetry lowering at low temperatures. **c**. Temperature dependence of Raman scattering spectra for $K_2WCl_6$ in the parallel geometry measurement. The split of $T_{2g}$ phonon mode near 170 cm$^{-1}$ is clearly observed.

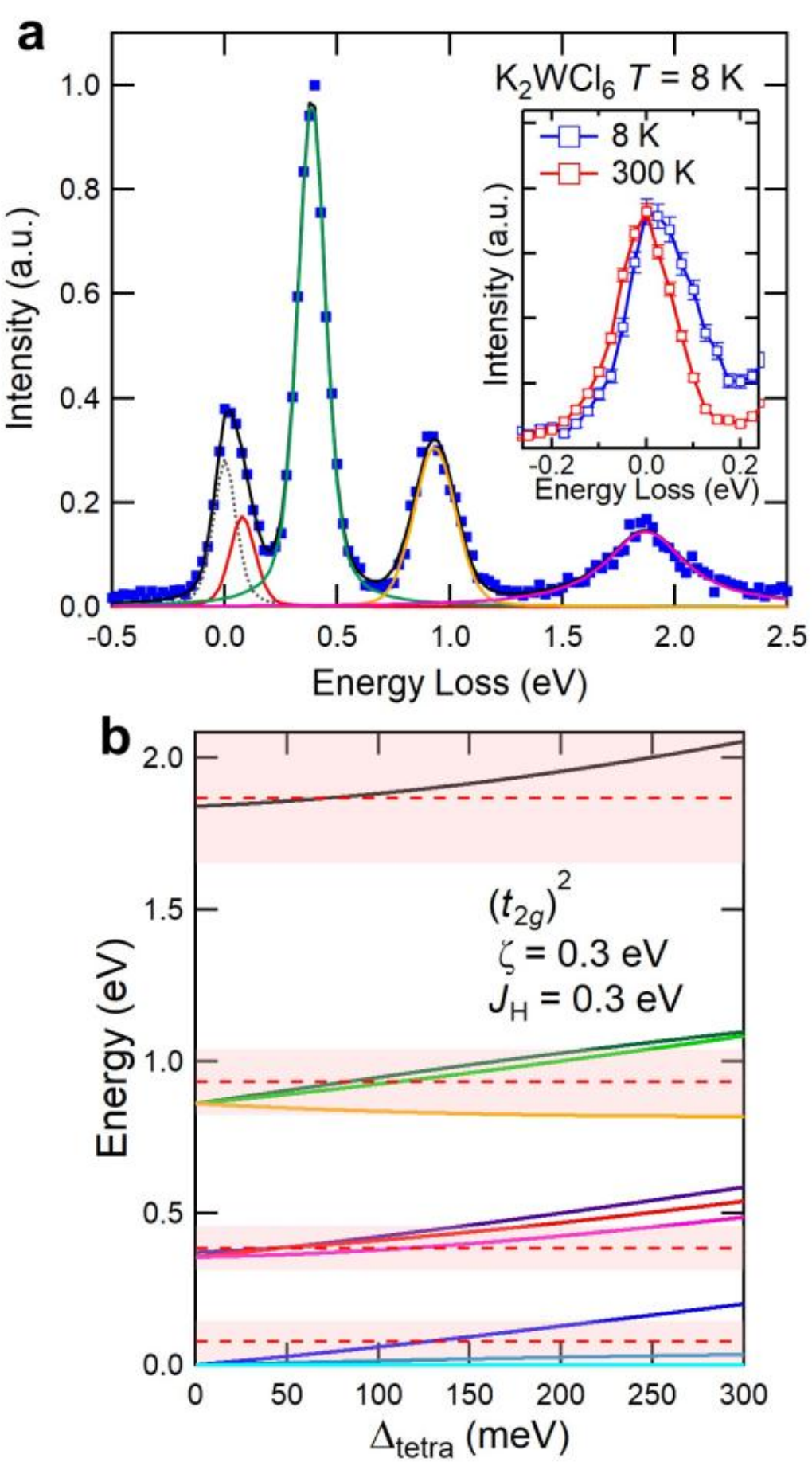

Fig. 8. Change of electronic structure in $K_2WCl_6$ associated with the structural transition. **a**. W $L_3$-edge RIXS spectra of $K_2WCl_6$ at 300 and 8 K. The energy of incident X-ray was 10.203 keV, and the spectra were collected at the L-point [**Q** = (6.5, 6.5, 6.5)] of the original cubic lattice. The black dotted line shows the contribution of elastic scattering. The inset shows a magnified view near the elastic scattering. Sizeable intensity is seen at low energies around 0.1 eV at 8 K, superposing on the elastic scattering. The contribution can be explained by a peak centered around 80 meV shown as the red line in the main panel. **b**. Calculated energy levels of spin-orbital multiplet as a function of elongational tetragonal crystal field $\Delta_{tetra}$ for a $(t_{2g})^2$ ion. In the calculation, both spin-orbit coupling $\zeta$ and Hund's coupling $J_H$ are fixed at 0.3 eV based on the result for $Rb_2WCl_6$. The $J = 2$ manifold splits into $J^z = 0$, ±1, and ±2 states by the tetragonal crystal field. The $J^z = 0$ singlet is the ground state, and the low-energy peak around 80 meV can be explained by the excitation between $J^z = 0$ and ±2 states with $\Delta_{tetra} \sim 130$ meV.

**Table 1 Structural parameters of $K_2WCl_6$ at 100 K refined by single-crystal X-ray diffraction.** The space group is *I*4/*mmm* (No. 139) and $Z = 2$. The lattice parameters are $a$ = 6.8001(16) Å and $c$ = 10.225(3) Å. The refinement indices are $R$ = 5.42%, and GOF = 1.149.

| Atom | Site | *g* | *x* | *y* | *z* | $U_{iso}$ (Å$^2$) |
|---|---|---|---|---|---|---|
| W | 2*a* | 1.0 | 0 | 0 | 0 | 0.0210(3) |
| K | 4*d* | 1.0 | 0.5 | 0 | 0.25 | 0.0261(6) |
| Cl1 | 4*e* | 1.0 | 0 | 0 | 0.2356(3) | 0.0244(6) |
| Cl1 | 16*l* | 0.5 | -0.2054(6) | -0.2806(6) | 0 | 0.0260(7) |

**Discussions**

A $5d^2$ transition-metal ion is expected to form a $J = 2$ state owing to the strong spin-orbit coupling. The $J = 2$ state has been proposed to display ordering of electronic multipoles such as charge quadrupole ordering or magnetic octupole ordering [8,9,18-22]. $5d^2$ double-perovskite oxides have been intensively studied in this regard, and the formation of magnetic octupole ordering was proposed in the double perovskite osmates [17] although their ground state has not yet been firmly established.

The tungsten chlorides $A_2WCl_6$ (A = Cs, Rb, and K) possess an electronic configuration and FCC sublattice of W ions similar to those of double-perovskite osmates. The spin-orbit-entangled $J = 2$ state was identified in $A_2WCl_6$ by RIXS, providing another platform for multipolar physics. $K_2WCl_6$ displays a structural transition at ~180 K, which lifts the degeneracy of $J = 2$ state and yields a singlet ground state. We argue that this transition is caused by structural instability owing to the size mismatch between the $K^+$ ion and $WCl_6$ octahedra, rather than by charge quadrupolar interactions.

By contrast, $Rb_2WCl_6$ and $Cs_2WCl_6$ did not show any signs of phase transitions down to low temperatures. Although the W-Cl bonds should have smaller degrees of hybridization and the magnitude of super-exchange interactions may be reduced compared to those of $5d$ oxides, the appearance of multipole ordering has been predicted at around 5.5 K for $Cs_2WCl_6$ from first-principle calculations [33], and the multiple interactions should not be negligibly small. Other factors like chemical disorder, frustration or both may prevent the formation of long-range multipolar ordering.

The structural investigation at low temperatures on $Rb_2WCl_6$ revealed signatures of local structural distortion despite the absence of a global structural transition. The (400) Bragg reflection in the PXRD measurement was broadened upon cooling, while the (111) peak did not show visible broadening. This suggests a local tetragonal distortion, possibly associated with the distortion of the $WCl_6$ octahedra. This distortion is unlikely to be caused by steric instability as in $K_2WCl_6$, because such distortions have not been observed in other $5d$ FCC chlorides with larger $Rb^+$ ions such as $Rb_2TaCl_6$ and $Rb_2ReCl_6$ [40, 41]. Instead, we propose that the weak structural anomaly is driven by incipient quadrupole ordering. The antiferro-type quadrupole interaction is frustrated on the FCC lattice [21], and the presence of disorder, such as oxygen incorporation [27] or A-site deficiency inferred from the ACl impurities (Fig. S1), may impede the formation of long-range quadrupolar ordering. Considering the broadening of phonon peak in the Raman scattering (Fig. 5**a**), that of PXRD Bragg reflections (Fig. 6**b**), and the change of temperature dependence of the lattice parameter (the inset of Fig. 6**a**), local distortions may set in around 100 K and gradually develop on cooling further. By scrutinizing the heat capacity data (Fig. 2**b**), the $C(T)$ of $A_2WCl_6$ [A = Rb, Cs] is slightly larger than that of $K_2WCl_6$ with quenched electronic degrees of freedom below ~100 K, which might reflect the gradual release of electronic entropy, although quantitative evaluation is difficult because of the different crystal structures.

The presence of weak structural distortion in $A_2WCl_6$ [A = Rb, Cs] implies that quadrupolar coupling prevails over the ferromagnetic octupolar interaction which is considered stronger in the Os double perovskites [17-19]. If the latter interaction is dominant, it is not frustrated and would form long-range octupolar ordering in spite of the presence of disorder, although the transition temperature may be reduced [42]. Compared to the Os double-perovskite oxides, the 10Dq of $A_2WCl_6$ [A = Rb, Cs] is smaller, and the $E_g$-$T_{2g}$ gap, which is in proportion to ~$\zeta^2$/10Dq, would be enhanced. This may reduce the ferro-octupolar coupling via the excited $T_{2g}$ state [18], making the octupolar coupling weaker than the quadrupolar one. In addition, coupling with phonons may strongly renormalize the magnitude of multipolar interactions [21,43,44]. The coupling of the spin-orbital state with local lattice deformation, i.e. dynamic Jahn-Teller effect, has been discussed to augment the strength of quadrupolar interactions [21,44]. The antifluorite-type structure of $A_2WCl_6$ has no intermediate cations between the $WCl_6$ octahedra; thus, the $WCl_6$ is fully isolated unlike in Os double-perovskites $Ba_2MOsO_6$ where non-magnetic cations, such as $Mg^{2+}$, $Ca^{2+}$, or $Zn^{2+}$, are bonded to the oxygen atoms of the $OsO_6$ octahedra. This structural feature may promote coupling with phonons more effectively than in double-perovskites, boosting quadrupolar interactions over octupolar one. The dynamic Jahn-Teller effect may also play a role in broadening the $E_g$-$T_{2g}$ excitation, making it difficult to observe the excitation by INS.

To summarize, the electronic, magnetic, and structural properties of antifluorite tungsten chloride $A_2WCl_6$ (A = Cs, Rb, and K) were investigated. Despite the five-fold $J$ = 2 state, $Rb_2WCl_6$ and $Cs_2WCl_6$ showed no clear signs of phase transition down to the lowest accessible temperature. A close inspection of the low-temperature structure points to the presence of subtle structural distortions. It is tempting to infer that such a structural instability is associated with the quadrupolar component of the $J$ = 2 state. We suspect that the presence of chemical disorders impedes the occurrence of long-range quadrupolar ordering, leading to a non-magnetic quadrupole-glassy state. To elucidate the nature of the ground state, local probes such as extended X-ray fine structure (EXAFS) measurements, pair-distribution function (PDF) analysis, or nuclear quadrupole resonance may provide decisive information on the low-temperature crystal structure. Alternatively, substantial improvement in sample quality may bring about a formation of long-range quadrupolar ordering. Finally, chemical modifications, such as 4$d$ counterparts $A_2MoCl_6$ or enhanced covalency like $A_2WBr_6$ and $A_2WI_6$, may give rise to distinct electronic ground states, which should be worthy of future investigations to further explore the rich multipolar physics of spin-orbit-coupled *d*-electrons.

## Methods

**Sample Preparation.** Polycrystalline samples of $A_2WCl_6$ (A = Cs, Rb, and K) were synthesized via a conventional solid-state reaction. Stoichiometric amount of ACl, W and $WCl_6$ powders were weighted and mixed in an Ar-filled glovebox. The mixture was pelletized and subsequently sealed in an evacuated quartz tube. The tube was heated at 650 °C for 72 hours. Single crystals of $Rb_2WCl_6$ and $K_2WCl_6$ were grown via a chemical vapor transport reaction using polycrystalline powder.

**Structural analysis.** The powder and single-crystal samples were characterized using X-ray diffraction. The powder X-ray diffraction measurements at room temperature were performed with Bruker D2 Phaser (Cu K$\alpha$ radiation), and the measurements at low temperatures were conducted with Bruker D-8 Advance system using Cu K$\alpha$1 radiation from primary Ge (111) Johannson monochromator. Oxford Cryosystem Phenix was used to change the temperature of the sample. Single-crystal X-ray diffraction data were collected at 298 and 100 K using a SMART APEX-I CCD X-ray diffractometer (Bruker AXS, Karlsruhe, Germany) equipped with a Cryostream 700 Plus cooling device (Oxford Cryo-systems, Oxford, U.K.). Data collection and reduction were carried out using the Bruker Suite software package.

**Magnetic and thermodynamic measurements.** The magnetic and thermodynamic properties were investigated down to 2 K using commercial equipment (Quantum Design MPMS3 and PPMS). The contributions from the core diamagnetic susceptibilities [45] have already been subtracted in the data show in Fig. 2**a**. The heat capacity measurements on the $Rb_2WCl_6$ single crystals at ultra-low temperatures were performed using a custom-built apparatus equipped with a dilution fridge [46].

**Resonant inelastic X-ray scattering.** The W $L_3$-edge RIXS measurements were carried out at BL11XU of SPring-8. Single crystals of $Rb_2WCl_6$ and $K_2WCl_6$ were used for the measurements, whereas a polycrystalline sample was used for $Cs_2WCl_6$. The energy of the incident X-ray was tuned to 10.203 keV which corresponds to the $2p_{3/2}$ to $5d$ $t_{2g}$ excitation energy of the $W^{4+}$ ion. The incident X-rays were monochromatized by a Si(444) channel-cut monochromator, and the scattered X-rays from the sample were analyzed by Si(555) diced and spherically-bend monochromator. The total energy resolution determined by the measurement of Kapton tape was about 120 meV.

The Re $L_3$-edge RIXS measurements shown in Supplementary Information were performed at the same beamline. A polycrystalline pellet of $Sr_2YReO_6$ was used in this experiment. The incident X-rays were monochromatized by a 4-bounced Si(400) asymmetric monochromator, and the scattered X-rays were analyzed by a Si(119) diced and spherically-bend analyzer. The energy of the incident X-ray was tuned to 10.536 keV. The total energy resolution was ~115 meV.

**Raman spectroscopy.** Raman scattering measurements were performed on single crystals of $Rb_2WCl_6$ and $K_2WCl_6$ in the full back-scattering geometry. The incident polarization was in an arbitrary direction within the (111) surface of single crystals. The measurements were performed using a 632.8 nm He-Ne laser source and a LabRAM HR800 spectrometer (HORIBA). The spectrum at each temperature and geometry was collected with a counting time of 120 seconds and 7 repetitions.

**Muon-spin relaxation.** Muon-spin relaxation measurements were conducted on powder samples of $A_2WCl_6$ on ARGUS at ISIS Neutron and Muon Source [47]. The measurements were performed at zero field and with a longitudinal field of 100 Oe. All the collected data were analyzed using the WIMDA software [48].

**Inelastic neutron scattering.** Time-of-flight inelastic neutron scattering was carried out using the thermal spectrometer MAPS at the ISIS Neutron and Muon source [49]. A 6.7 g polycrystalline sample was used for the experiment. Measurements were conducted with the incident energy $E_i$ = 100 meV, 70 meV, 50 meV and 20 meV and chopper frequency of 400 Hz, 350 Hz, 350 Hz and 200 Hz, respectively.

**Acknowledgement**

We are grateful to G. Khaliullin, N. Iwahara, G. Jackeli, and K. M. Kojima for invaluable discussions. The synchrotron radiation experiment was performed at the BL11XU of SPring-8 with the approval of the Japan Synchrotron Radiation Research Institute (JASRI) (Proposal No. 2019B3555 and 2020A3555). We thank the RIKEN-RAL muon facilities for the allocation of beamtime on ARGUS at ISIS Neuron and Muon Source (Proposal No. RB2070012). We acknowledge the provision of beamtimes on MAPS (Proposal No. RB2190103). This work was partly supported by Alexander von Humboldt Foundation, and by Izumi Science and Technology Foundation.